\def\year{2022}\relax
\documentclass[letterpaper]{article} 

\usepackage{aaai22}  
\usepackage{times}  
\usepackage{helvet}  
\usepackage{courier}  
\usepackage[hyphens]{url}  
\usepackage{graphicx} 
\usepackage{natbib}  
\usepackage{caption} 
\DeclareCaptionStyle{ruled}{labelfont=normalfont,labelsep=colon,strut=off} 
\usepackage{algorithm}
\usepackage{algorithmic}

\usepackage{newfloat}
\usepackage{listings}
\floatstyle{ruled}
\newfloat{listing}{tb}{lst}{}
\floatname{listing}{Listing}
\title{How Decentralization Affects User Agency on Social Platforms}
\author{
    Aditya Surve\textsuperscript{1}\thanks{Equal contribution.},
    Aneesh Shamraj\textsuperscript{1}\footnotemark[1],
    Swapneel Mehta\textsuperscript{1}
}
\affiliations {
    \textsuperscript{\rm 1} 
    SimPPL, ssaditya2002@gmail.com
}
\usepackage{bibentry}

\begin{document}

\maketitle

\begin{abstract}
Mainstream social media platforms function as "walled garden" ecosystems that restrict user agency, control, and data portability.
They have demonstrated a lack of transparency that contributes to a multitude of online harms. 
Our research investigates how decentralization might present promise as an alternative model to walled garden platforms. Specifically, we describe the user-driven content moderation through blocks as an expression of agency on Bluesky, a decentralized social platform. 
We examine the impact of providing users with more granular control over their online experiences, including what they post, who can see it, and whose content they are exposed to. 
We describe the patterns identified in user-driven content moderation and suggest directions for further research.
\end{abstract}

\section{Introduction}

Mainstream social platforms, such as Facebook, Twitter, and Instagram, play a crucial role in driving a large volume of online interactions \cite{Jarad2014MarketingOS,ElenaIulia2018THEIO}. 
They also significantly impact consumer behavior, with a brand's presence on social media influencing future purchases \cite{ElenaIulia2018THEIO}. 
Furthermore, social networks cater to diverse needs for interaction, including expressing opinions, sharing information, following brands, and consuming content \cite{article}.
\par
However, the current social web is characterized by a "walled garden" structure, where major social media platforms operate as isolated data silos, trapping user data and relationships within their proprietary ecosystems. This fragmentation not only hinders the free flow of information but also presents significant challenges to user privacy and control. Existing social media platforms often provide limited and rigid privacy settings, allowing users to only define simple mappings between predefined categories of personal information and authorized groups of contacts. Moreover, these privacy preferences are typically confined to individual platforms, making it difficult for users to consistently apply their desired policies across their diverse online activities \cite{info13060280}.
\par
As the social media landscape continues to evolve, there is a growing need for more decentralized and user-centric approaches that empower individuals to have greater control over their personal data and online interactions. \cite{karger2010guarding}
The lack of transparency around algorithms that curate content is also mentioned as potentially influencing what users see. Algorithmic curation allows platforms to selectively show content to users based on opaque algorithms and undisclosed factors, potentially influencing what information users are exposed to and the choices they make. This curation happens without full transparency to users regarding the working of algorithms, with no knowledge of the criteria that determine what content is prioritized. In fact, platform recommender systems are often so complex they are uninterpretable even by the designers of these algorithms. By controlling what users see through algorithmic curation and making it hard to leave through data portability barriers, centralized platforms can reduce user autonomy and agency over their online experiences \cite{gramegna2018data}.
\par
These platforms utilize dark patterns to reduce user agency and maintain their control over the user experience. According to a study, centralized social media platforms like Facebook, Instagram, TikTok, and Twitter employ various dark patterns as explained by \cite{Mathur_2019} that can reduce user agency and control on the platform. Interface interference and visual interference are two examples of deceptive design which privilege certain interface elements over others to coerce users into making particular choices.This includes making options for sharing personal data more prominent and harder to avoid. Deceptive interface designs and default settings nudge users towards choices that benefit the platform by exposing more user data, while reducing the user's control over their privacy and data sharing preferences. The lack of data portability makes it difficult for users to export their data and content from one platform to switch to another.
\noindent
\par
Bluesky is a decentralized social network created by Jack Dorsey, the former Twitter CEO. The platform was developed in 2019 as a decentralized, social network built on an open-source protocol, the AT Protocol. \cite{kleppmann2024bluesky} Bluesky is designed to not be controlled by a single company and operates on an open-source model, promoting transparency and community involvement. The platform is very similar to Twitter, allowing users to create profiles, follow other users, and create posts with a maximum of 300 characters. The platform began as an invite-only app, requiring an invite code from a current user to join the Bluesky waitlist, but has since opened up to all users. 
\section{Existing research and studies}
There are several research works on the decentralised social media platforms. \cite{Zignani2018FollowT} presents a dataset containing both the network of the "follow" relationships and its growth in terms of new connections and users, all which were obtained by mining the decentralized online social network named Mastodon. The dataset is combined with usage statistics and meta-data about the servers comprising the platform's architecture, which are called instances. The paper also analyzes the overall structure of the Mastodon social network, focusing on its diversity w.r.t. other commercial microblogging platforms such as Twitter.

 \cite{la2021understanding} aims at pushing forward our understanding of the Fediverse by leveraging the primary role of Mastodon therein. The authors build an up-to-date and highly representative dataset of Mastodon and define a network model over Mastodon instances to investigate the structural features of the Mastodon network of instances from a macroscopic as well as a mesoscopic perspective, the backbone of the network, and the growth of Mastodon.

\cite{masto2} studies the Twitter migration to Mastodon following Elon Musk’s acquisition and analyzes the networks of social links between these 75K users on both Twitter and Mastodon. The authors found around 75K valid Mastodon handles that were associated with active Mastodon accounts and analyzed the differences between the two networks in terms of density, average degree, transitivity, and the presence of small disconnected components.

\cite{masto} investigates how a decentralized architecture based on distributed servers impacts the structure of the users' social relationships in Mastodon. The authors found that the decentralized architecture of Mastodon leads to a more sparse network compared to centralized social media platforms, and that users tend to form smaller and more cohesive communities.
These research works provide valuable insights into the structure, evolution, and behavior of users in decentralized social media platforms like Mastodon.

The decentralised social media platform Diaspora has been a subject of academic interest, with various studies exploring its impact and dynamics. One notable research work is \cite{andersson} which explores the relationship between new media and transnational social fields, shedding light on the implications of digital diasporas.

\cite{guidi} discusses Steem and Hive as examples of blockchain-based social media platforms and their potential to offer control to users over their data and content, as well as the ability to manage monetization.

These studies provide insights into the technical, economic, and security aspects of decentralized social media platforms, highlighting their potential to offer control to users, promote high-quality content, and create a new paradigm for online social networks.

\section{Bluesky's Role in Addressing Centralized Social Media Challenges}
Centralized social media platforms encounter a plethora of challenges in today's digital landscape. They have started making profitability and business a priority over the benefits they provide to the users to the point that users have been unhappy with them\cite{ruwe2023difference}
Privacy concerns loom large as users grow increasingly wary of how their data is collected, stored, and utilized by these platforms, following high-profile data breaches and controversies. 
Content moderation poses a big challenge, with platforms struggling to strike a balance between freedom of expression and curtailing harmful content, often leading to debates around censorship and misinformation.\cite{kozyreva2023resolving}
Fake accounts and bots continue to plague platforms, perpetuating misinformation and manipulating user engagement metrics, as evidenced by Twitter's efforts to remove millions of fake accounts \cite{thomas2011suspended}
\par
Bluesky is built on a decentralized architecture, where multiple providers can offer interoperable services for different components of the system.
This decentralization aims to avoid the concentration of power and control under a single entity, as seen in centralized platforms \cite{bailey2022interoperability}
Bluesky allows users to easily switch between different providers for their personal data server (PDS), feed generators, and moderation services.
Users have more control over the content they see, as they can choose the moderation services and feed algorithms they use 
Bluesky uses decentralized identifiers (DIDs) and DNS domain names as user handles, allowing users to change their PDS provider without changing their identity.
This addresses the lock-in issues seen in federated systems like Mastodon, where changing servers often requires changing usernames.
Bluesky's indexing infrastructure, comprising the Relay and App View services, enables a global view of the network without overburdening individual providers.
This approach aims to provide a user experience comparable to centralized platforms while maintaining the decentralized nature of the system.\cite{kleppmann2024bluesky}

\section{Data Description}

\graphicspath{ {./images/} }
\begin{figure*}
\includegraphics[trim={0 0 0 1.85cm},clip]
{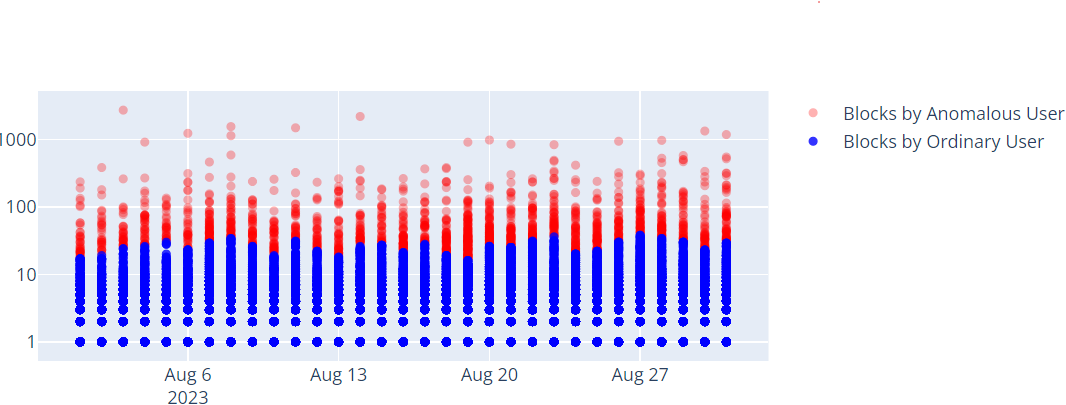}
\caption{Distribution of Anomalous Blocking Behavior (top percentile each day)}
\label{fig:op}
\end{figure*}
Due to the decentralized nature of Bluesky, almost all of the data is accessible for collection through the AT Protocol\footnote{\texttt{\url{https://atproto.com/}}}. 
All of the data of a particular user is stored in a single place known as a repository. 
Repositories are unique to each user on the platform and free to access through third-parties. 
This allows the user to migrate to any other platform based on the AT Protocol and migrate all their data to another platform without any data loss. 
Every user is identified by a unique id saved as their ‘DID’, referring to the term 'decentralized identifier'. 
We use this identifier to extract the data of each user relevant to our study. 
First we obtain all the ‘DID’s by a ‘GET’ request on the ‘https://bsky.network/xrpc/com.atproto.sync.listRepos’ route. 
Then we extract the data of that user. 
The dataset for Bluesky data which we have collected consists of the following tables: blocks, follows, users, reposts, posts, tags, links and mentions. 
The description of these tables are as follows:

\textbf{Blocks:} 
A row in this table represents a social block. This shows us if a particular user has blocked another user. This is done by fetching data by a ‘GET’ request at '\url{https://bsky.social/xrpc/com.atproto.repo.listRecords?collection=app.bsky.graph.block&repo={did}}’

\textbf{Follows:}
A row in this table represents a social follow. This shows us if a particular user has followed another user. This is done by fetching data by a ‘GET’ request at ‘\url{https://bsky.social/xrpc/com.atproto.repo.listRecords?collection=app.bsky.graph.follow&repo={did}}’

\textbf{Users:}
A row in this table represents all the details of a particular user such as user handle, url of the user profile, name of the user as well as the description of the account set by the user. This is done by fetching data by a ‘GET’ request at ‘\url{https://bsky.social/xrpc/com.atproto.repo.describeRepo?repo={did}}’ and ‘\url{https://bsky.social/xrpc/com.atproto.repo.listRecords?repo={did}&collection=app.bsky.actor.profile}’ to get all the details.

\textbf{Reposts:} 
A row in this table shows the data regarding a single repost made by the user. This shows which user reposted a particular post. This is done by fetching data by a ‘GET’ request at ‘\url{https://bsky.social/xrpc/com.atproto.repo.listRecords?collection=app.bsky.feed.repost&repo={did}}’

\textbf{Posts:}
A row in this table shows the data regarding a single repost made by the user. This shows which user reposted a particular post. This is done by fetching data by a ‘GET’ request at ‘\url{https://bsky.social/xrpc/com.atproto.repo.listRecords?collection=app.bsky.feed.post&repo={did}}’

\textbf{Tags:}
This is a sub table created by using the data fetched while creating the posts table. A row in this table represents a particular hashtag used by a user in a particular post.

\textbf{Links:}
This is a sub table created by using the data fetched while creating the posts table. A row in this table represents a particular link to any other website used by a user in a particular post.

\textbf{Mentions:}
This is a sub table created by using the data fetched while creating the posts table. A row in this table represents a user mentioned by another user in a particular post.

All of the rows of all tables mentioned above have timestamps in the form of date and time columns. This data is also returned by the AT Proto API. All of these API endpoints return a set number of entries like a maximum of 1000 for the ‘\url{https://bsky.network/xrpc/com.atproto.sync.listRepos}’ route and a maximum of 100 for ‘\url{https://bsky.social/xrpc/com.atproto.repo.listRecords}’ route. A ‘cursor’ value is returned along with those entries which should be added to the next get request to get the records after the current records. In this way all the data is extracted.

\section{Findings}
In this study, we leveraged the Bluesky API to procure comprehensive data regarding user blocks throughout the month of August. This dataset encompassed crucial information, including the identity of users involved in blocking interactions, the precise dates, and timestamps of these events. Our primary objective was to discern any anomalous patterns exhibited by users in their blocking behaviors during this temporal scope. Employing statistical techniques, we computed the z-scores for the frequency of blocks per user on a daily basis. These z-scores served as pivotal metrics, enabling us to pinpoint outliers by quantifying the deviation of block counts from the mean in terms of standard deviations. Specifically, users whose z-scores surpassed the 99th percentile threshold for a given day were classified as anomalous, while the remaining users were categorized as regular.
\par
Figure \ref{fig:op} generated from our analysis vividly illustrates the disparity between the blocking activities of anomalous users and those of their regular counterparts. Notably, the plot delineates two distinct categories of users based on their blocking behaviors: anomalous users are depicted by sporadic red markers, indicating instances where their block counts significantly exceeded the expected levels for a particular day, while ordinary users are represented by a dense cluster of blue markers, denoting their consistent and comparatively lower blocking activity. This visual contrast underscores the pronounced deviation exhibited by anomalous users from the norm. Through this comprehensive analysis, we have uncovered a notable subset of users whose behavior deviates markedly from the general population, thereby shedding light on potential anomalies within the observed dataset.
\section{Conclusion and Future work}
Bluesky aims to address several challenges faced by centralized platforms, such as privacy concerns, content moderation issues, and the proliferation of fake accounts and bots.
Our analysis of user blocking behavior on Bluesky revealed distinct patterns, with a subset of users exhibiting anomalous activity that significantly deviated from the norm. This finding highlights the potential for user-driven content moderation to identify and address problematic behavior on decentralized platforms effectively.
While Bluesky is still in its early stages, the platform's emphasis on transparency, user agency, and data portability holds promise for reshaping the social media landscape.
\par
While this study offers valuable insights into user behavior and content moderation on the decentralized social platform Bluesky, several avenues for future research emerge. Conducting longitudinal analyses could track the evolution of user dynamics and content moderation over extended periods. Comparing user behavior and network structures across different decentralized platforms like Mastodon and Diaspora could yield insights into their respective strengths and weaknesses. Evaluating the scalability and performance of Bluesky's decentralized architecture under increasing user loads will be crucial. User studies could inform the development of more user-friendly interfaces and drive broader adoption. Exploring different algorithms and policies for user-driven content moderation, as well as AI-assisted moderation, could lead to more effective practices. Investigating privacy, security implications, and potential vulnerabilities in decentralized systems is essential. Finally, researching sustainable economic models and incentive structures involving tokenization and micropayments could foster long-term viability and adoption of decentralized social media platforms. 

\bibliography{aaai22}

\begin{thebibliography}{18}
\providecommand{\natexlab}[1]{#1}

\bibitem[{Andersson(2019)}]{andersson}
Andersson, K.~B. 2019.
\newblock Digital diasporas: An overview of the research areas of migration and new media through a narrative literature review.
\newblock \emph{Human Technology}, 15: 142--180.

\bibitem[{Bailey and Misra(2022)}]{bailey2022interoperability}
Bailey, R.; and Misra, P. 2022.
\newblock Interoperability of social media: an appraisal of the regulatory and technical ecosystem.
\newblock \emph{Available at SSRN 4095312}.

\bibitem[{Elena-Iulia(2018)}]{ElenaIulia2018THEIO}
Elena-Iulia, V. 2018.
\newblock THE IMPORTANCE OF SOCIAL MEDIA.
\newblock \emph{Annals - Economy Series}, 6: 80--91.

\bibitem[{Gramegna(2018)}]{gramegna2018data}
Gramegna, A. 2018.
\newblock Data-Gathering, Governance, and Algorithms: How Accountable and Transparent Practices Can Mitigate Algorithmic Threats.

\bibitem[{Guidi(2021)}]{guidi}
Guidi, B. 2021.
\newblock An Overview of Blockchain Online Social Media from the Technical Point of View.
\newblock \emph{Applied Sciences}, 11: 9880.

\bibitem[{Jarad(2014)}]{Jarad2014MarketingOS}
Jarad, G.~A. 2014.
\newblock Marketing Over Social Media Networks.
\newblock \emph{European Journal of Business and Management}, 6: 114--117.

\bibitem[{Kaplan and Haenlein(2010)}]{article}
Kaplan, A.; and Haenlein, M. 2010.
\newblock Users of the World, Unite! The Challenges and Opportunities of Social Media.
\newblock \emph{Business Horizons}, 53: 59--68.

\bibitem[{K{\"a}rger and Siberski(2010)}]{karger2010guarding}
K{\"a}rger, P.; and Siberski, W. 2010.
\newblock Guarding a walled garden—semantic privacy preferences for the social web.
\newblock In \emph{Extended Semantic Web Conference}, 151--165. Springer.

\bibitem[{Kleppmann et~al.(2024)Kleppmann, Frazee, Gold, Graber, Holmgren, Ivy, Johnson, Newbold, and Volpert}]{kleppmann2024bluesky}
Kleppmann, M.; Frazee, P.; Gold, J.; Graber, J.; Holmgren, D.; Ivy, D.; Johnson, J.; Newbold, B.; and Volpert, J. 2024.
\newblock Bluesky and the AT Protocol: Usable Decentralized Social Media.
\newblock \emph{arXiv preprint arXiv:2402.03239}.

\bibitem[{Kozyreva et~al.(2023)Kozyreva, Herzog, Lewandowsky, Hertwig, Lorenz-Spreen, Leiser, and Reifler}]{kozyreva2023resolving}
Kozyreva, A.; Herzog, S.~M.; Lewandowsky, S.; Hertwig, R.; Lorenz-Spreen, P.; Leiser, M.; and Reifler, J. 2023.
\newblock Resolving content moderation dilemmas between free speech and harmful misinformation.
\newblock \emph{Proceedings of the National Academy of Sciences}, 120(7): e2210666120.

\bibitem[{La~Cava, Aiello, and Tagarelli(2023)}]{masto2}
La~Cava, L.; Aiello, L.; and Tagarelli, A. 2023.
\newblock Drivers of social influence in the Twitter migration to Mastodon.
\newblock \emph{Scientific Reports}, 13.

\bibitem[{La~Cava, Greco, and Tagarelli(2021)}]{la2021understanding}
La~Cava, L.; Greco, S.; and Tagarelli, A. 2021.
\newblock Understanding the growth of the Fediverse through the lens of Mastodon.
\newblock \emph{Applied network science}, 6: 1--35.

\bibitem[{Liu et~al.(2022)Liu, Tse, Kwok, and Chiu}]{info13060280}
Liu, Y.; Tse, W.~K.; Kwok, P.~Y.; and Chiu, Y.~H. 2022.
\newblock Impact of Social Media Behavior on Privacy Information Security Based on Analytic Hierarchy Process.
\newblock \emph{Information}, 13(6).

\bibitem[{Mathur et~al.(2019)Mathur, Acar, Friedman, Lucherini, Mayer, Chetty, and Narayanan}]{Mathur_2019}
Mathur, A.; Acar, G.; Friedman, M.~J.; Lucherini, E.; Mayer, J.; Chetty, M.; and Narayanan, A. 2019.
\newblock Dark Patterns at Scale: Findings from a Crawl of 11K Shopping Websites.
\newblock \emph{Proceedings of the ACM on Human-Computer Interaction}, 3(CSCW): 1–32.

\bibitem[{Ruwe(2023)}]{ruwe2023difference}
Ruwe, M. 2023.
\newblock \emph{The difference between centralized social media platforms and upcoming decentralized social media platforms: A comparative study}.
\newblock {B.S.} thesis, University of Twente.

\bibitem[{Thomas et~al.(2011)Thomas, Grier, Song, and Paxson}]{thomas2011suspended}
Thomas, K.; Grier, C.; Song, D.; and Paxson, V. 2011.
\newblock Suspended accounts in retrospect: an analysis of twitter spam.
\newblock In \emph{Proceedings of the 2011 ACM SIGCOMM conference on Internet measurement conference}, 243--258.

\bibitem[{Zignani, Gaito, and Rossi(2018)}]{Zignani2018FollowT}
Zignani, M.; Gaito, S.; and Rossi, G.~P. 2018.
\newblock Follow the "Mastodon": Structure and Evolution of a Decentralized Online Social Network.
\newblock In \emph{International Conference on Web and Social Media}.

\bibitem[{Zignani et~al.(2019)Zignani, Quadri, Gaito, Cherifi, and Rossi}]{masto}
Zignani, M.; Quadri, C.; Gaito, S.; Cherifi, H.; and Rossi, G.~P. 2019.
\newblock The Footprints of a “Mastodon”: How a Decentralized Architecture Influences Online Social Relationships.
\newblock In \emph{IEEE INFOCOM 2019 - IEEE Conference on Computer Communications Workshops (INFOCOM WKSHPS)}, 472--477.

\end{thebibliography}

\end{document}